\documentclass[twocolumn,
superscriptaddress,
amsmath,
amssymb,
aps,
prx]
{revtex4-2}

\usepackage{amsmath,amssymb,amsfonts,dsfont,xspace,graphicx,relsize,bm,mathtools,xcolor,amsthm,soul}
\usepackage{graphicx}
\usepackage{dcolumn}
\usepackage{xcolor}
\usepackage{bbm}
\usepackage{physics}
\usepackage{ulem}
\usepackage{hyperref}
\hypersetup{
	colorlinks = true,
	urlcolor   = blue,
	linkcolor  = blue,
	citecolor  = magenta
}
\usepackage{algorithm}
\usepackage{algpseudocode}
\usepackage{float}
\usepackage[shortlabels]{enumitem}
\usepackage{subfigure}
\usepackage{orcidlink}

\begin{document}

\title{Noisy dynamics of confined quantum walks on a chip}

\author{L. Sansoni\orcidlink{0000-0002-4445-1036}}
\email{linda.sansoni@enea.it}
\affiliation{ENEA - Nuclear Department, Via E. Fermi 45, 00100 Frascati, Italy}

\author{E. Stefanutti\orcidlink{0000-0002-2833-7049}}
\affiliation{ENEA - Nuclear Department, Via E. Fermi 45, 00100 Frascati, Italy}
\author{C. Benedetti\orcidlink{0000-0002-8112-4907}}
\affiliation{Dipartimento di Fisica ``Aldo Pontremoli'', Universit\`a degli Studi di Milano, via~Celoria~16, I-20133 Milan, Italy}
\author{I. Gianani\orcidlink{0000-0002-0674-767X}}
\affiliation{Dipartimento di Scienze, Universit\'a degli Studi Roma Tre, Via della Vasca Navale 84, 00146 Rome, Italy}
\author{C. Taballione\orcidlink{0000-0001-8464-4417}}
\affiliation{QuiX Quantum B.V., 7521 AN Enschede, The Netherlands}
\author{A. Toor}
\affiliation{QuiX Quantum B.V., 7521 AN Enschede, The Netherlands}
\author{L. Herrera}
\affiliation{QuiX Quantum B.V., 7521 AN Enschede, The Netherlands}
\author{M. Pistilli}
\affiliation{ENEA - Nuclear Department, Via E. Fermi 45, 00100 Frascati, Italy}
\author{S. Santoro\orcidlink{0000-0002-9588-3816}}
\affiliation{ENEA - Nuclear Department, Via E. Fermi 45, 00100 Frascati, Italy}
\author{M. Barbieri\orcidlink{0000-0003-2057-9104}}
\affiliation{Dipartimento di Scienze, Universit\'a degli Studi Roma Tre, Via della Vasca Navale 84, 00146 Rome, Italy}
\affiliation{Istituto Nazionale di Ottica - CNR, L.go Enrico Fermi 6, 50126, Florence, Italy}
\affiliation{INFN Sezione Roma Tre, Via della Vasca Navale, 84, 00146 Rome, Italy}
\author{A. Chiuri\orcidlink{0000-0001-9733-0740}}
\affiliation{ENEA - Nuclear Department, Via E. Fermi 45, 00100 Frascati, Italy}

\begin{abstract}
Quantum walks represent an excellent testbed for investigating the interplay between unitary coherent and incoherent dissipative processes. Thanks to photonic quantum interferometers of considerable size, experimental studies could be performed, devoted to investigating the consequences of different sorts of realistic noise in these systems. In this work we employ a 20x20 on-chip multimode interferometer to introduce another key aspect in the problem: the presence of edges in the walker lattice, enforcing a confined evolution. We show how noise can disrupt translational symmetry and reshape interference patterns. The non trivial probability distributions obtained along the temporal evolution of the system demonstrate how speed up effects, localization and coherent oscillations are pillar concepts to be fully characterized and understood when applied in realistic quantum dynamics.
\end{abstract}

\maketitle

\section{Introduction\label{sec:intro1}}

Random walks are employed as a model to describe stochastic processes exploring a configuration space in the absence of deterministic rules. This unavailability may either be intrinsic to the system, or simply reflect incomplete  knowledge of underlying mechanisms. When moving to the quantum realm, quantum walks (QWs) are used in order to describe a similar unitary evolution of a quantum particle. In this sense, these preserve some of the features of  classical random walks, while also creating and preserving quantum coherences during the exploration process \cite{kempe003,venegas12}.
Quantum walks are classified depending on whether the evolution occurs in a continuous time (CT)~\cite{farhi1998,farhi98,frigerio21}, or  in discrete-time (DT) steps \cite{aharonov1993}. CTQWs describe the continuous evolution of a quantum particle over a discrete set of positions, governed by a Hamiltonian that encodes the connectivity of the underlying graph. In a DTQW, instead, the quantum walker  is equipped with an internal degree of freedom, referred to as quantum coin, and evolves in discrete time steps  over  discrete positions.   
The resulting dynamics  depend sensitively on both the initial state and the specific choice for the  evolution operator \cite{Tregenna03,kiss08}.

A characteristic signature of the behaviour of a quantum walker, both in CT and DT, is its ballistic propagation resulting from interference between different evolution paths \cite{ambainis01, Schreiber10,xue13}.  In contrast, classical random walks exhibit  a diffusive spreading that, for instance, leads  to a Gaussian probability distribution on an infinite line and to a stationary distribution on finite graphs.  In addition, quantum features such as superposition and interference give rise to structured probability distributions and enable dynamics with no classical analogue.
These properties reveal their potential to provide  quantum advantages in tasks relevant for quantum information processing and quantum technologies.
This simple idea is indeed so powerful that QWs can become a universal model to describe a wide variety of phenomena, ranging from quantum computing \cite{Lovett2010,webb13} and quantum algorithms  \cite{ambainis03,shenvi03,childs04,aaronson05,Eisenberg2005}, to quantum transport \cite{mulken11,Chan_2023} and state transfer and routing \cite{kay11,Kurzynski2011,zhan14,ragazzi25}. They are also employed in quantum metrology protocols \cite{annabestani22,singh23,gianani23,benedetti24,cavazzoni24} and for the implementation of entangling measurements \cite{li21,yan23,sengupta25}.
These considerations have sparked  considerable efforts to implement quantum walks in a variety of physical platforms, such as trapped atoms \cite{Karski09,Genske13}, trapped ions \cite{Schmitz09,Zahringer10}, 
nuclear magnetic resonance (NMR) systems  \cite{Ryan05} and photonic quantum systems, in bulk optics \cite{Broome10,Kitagawa12}, waveguide structures \cite{Crespi13,Peruzzo10,Owens11,Perets08,sansoni12,Solntsev12,Poulios14,crespi2021,peruzzo23} and fiber loop networks \cite{Schreiber10,Schreiber11,Schreiber12}. 
 
The investigation of QW-based protocols typically starts by assuming precise control of the walker’s dynamics, including  initial state preparation,  position-dependent access and manipulation of its quantum state and control of the quantum coin. In realistic scenarios, however, noise and fabrication defects can significantly affect the quantum evolution, altering the distinctive features of the QW.  A variety of noise mechanisms have been investigated in the context of QWs, including decoherence in the coin \cite{brun03d,balnus06,tude22},  links fluctuations and percolation  \cite{oliveira06,kollar14,Benedetti18}, static disorder \cite{Yin2008,Schreiber11,Jackson12}, each influencing the walker’s quantum properties in distinct ways.
Noise can indeed suppress interference effects and degrade the coherent dynamics, inducing a quantum-to-classical transition \cite{brun03}. 
In some regimes, static and quasi-static disorder give rise to Anderson localization, where destructive interference halts the spreading of the walker altogether \cite{anderson58,Crespi13}.  
However, noise is not always detrimental. Under certain conditions, it can enhance transport efficiency or coherence revival, playing a constructive role in phenomena such as environment-assisted quantum transport \cite{Yin2008,mohs08jcp,Stefanak16,Derevyanko2018}.
These effects have been studied in both the single and the multi-particle QW, revealing the impact of environmental interactions and fabrication imperfections  on the quantum dynamics of the walker \cite{rigovacca16,siloi17,sun18}.

Previous investigations have for the most part considered cases in which the transverse size of the walk ({\it e.g.} the number of modes in a waveguide array) was sufficiently large that the longitudinal evolution (the length of the waveguides) hardly populated the furthest sites. Therefore, unrestricted propagation of QWs across their available configuration space has been studied and realised. 
Thanks to the availability of larger reconfigurable structures~\cite{Shadbolt12,Taballione2021,Hoch22,Maring24}, one can now include effects from the boundaries. Crucially, these impose non-trivial constraints on the walker dynamics: they not only break translational invariance but also modify interference patterns, alter hitting-time statistics and  affect  probability distributions \cite{volta05,agliari08,Bach04,kwek11,Li18}.

Therefore, confinement is a key ingredient for understanding the dynamical features of QWs in realistic architectures, where finite-size effects and boundary-induced reflections cannot be neglected.  To date, only limited experimental effort has been dedicated to observing QWs in such restricted geometries, all the more in the presence of noise. Most existing studies have addressed either the coherent propagation in bounded systems or the impact of noise in unbounded settings, but their interplay has remained largely unexplored.
Yet,  such small finite systems can serve as effective minimal models for mimicking energy transport in complex networks, such as those found in biological light-harvesting complexes  \cite{Biggerstaff16,mohs08jcp,ferracin19}. 

In this work we close this gap by experimentally investigating DTQW dynamics under both spatial confinement and controlled noise in a universal quantum computer. By jointly exploring boundary effects and decoherence, we assess how these two mechanisms compete  in  creating new interference patterns and propagation properties. Our approach is based on arbitrary $20\times20$ transformations suitably designed to simulate the desired dynamic over several modes of a photonic chip and the steps of the linear chain have been exploited to mimic a time-dependent evolution. Thanks to the flexibility of the employed universal physical platform we could inspect a variety of configurations, and different kinds of noisy dynamics.  
This perspective allows us to analyze regimes that are relevant for realistic networked quantum devices, where understanding and accurately characterizing the interplay between boundary-induced effects and noise is essential for realizing controllable and reliable quantum operations.

\begin{figure}[h]
    \centering
    \includegraphics[width=0.9\columnwidth]{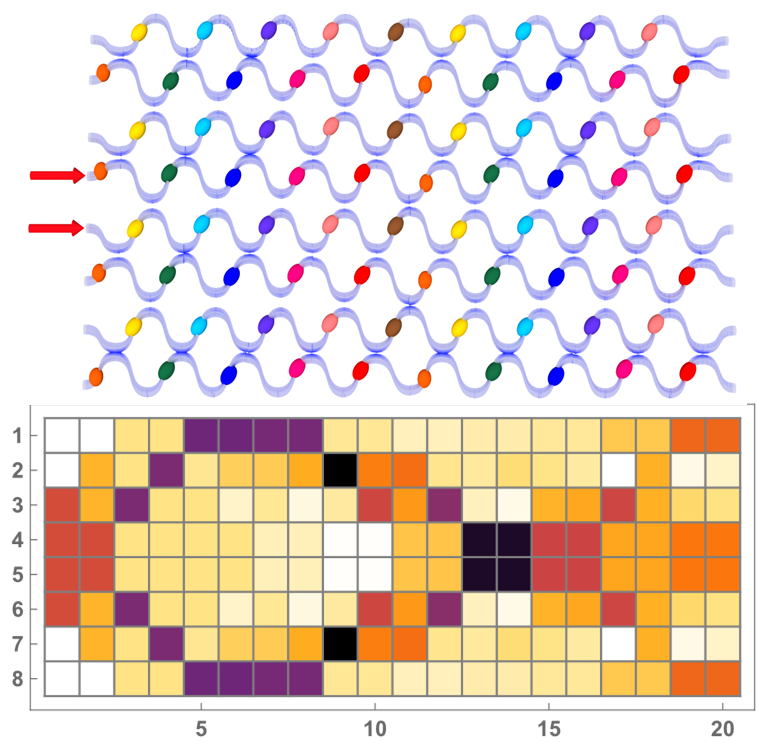}
    \caption{Top: scheme of a confined 8-site QW in the photonic processor built as a cascade of Mach-Zehender interferometers. Light is injected from the two central input waveguides (red arrows). Colored ovals on the waveguides represent phase shifters. Bottom: probability distribution of a confined 8-site Hadamard QW up to 20 steps.} 
    \label{fig:cascade}
\end{figure}

\section{Background} 

A numerical study of the propagation dynamics of photonic walkers is carried out in both ordered and disordered environments, focusing on how different types and strengths of disorder affect the spreading of the photonic wavefunction.  

The description of our DTQW considers the evolution of a walker as a result of two operations: first a coin flip, and then a spatial shift to the left or to the right, conditioned on the coin outcome. Therefore, the QW entails a joint Hilbert space in the form
$$\mathcal{H} = \mathcal{H}_C \otimes \mathcal{H}_S. $$
 $\mathcal{H}_S$ is the position (or shift) space representing the external degree of freedom, spanned by the orthonormal basis states $\ket{x}$ ($x \in \mathds{Z}$) giving the position of the walker along the finite discrete line. $\mathcal{H}_C$ is the coin space, which is a two-dimensional space spanned by vectors $\ket{L}$ and $\ket{R}$, corresponding to leftward and rightward propagation, respectively. 
 The quantum walker  starts at the origin and tosses a two-state quantum coin (i.e. a qubit). 
 
 As a result, a generic state of the system can be written as:
\begin{equation}
    \ket{\Psi} = \sum_{x \in \mathds{Z}} \ket{\psi(x)} \otimes \ket{x}
\end{equation}
where $\ket{\psi(x)} = \alpha(x)\ket{L}+\beta(x)\ket{R}$ (with $\alpha$ and $\beta$ complex probability amplitudes such that $|\alpha|^2+|\beta|^2=1$) is a superposition of the two outcomes for the coin. At the following time step, the walker at site $x$ can thus move to the left with probability $|\alpha|^2$ and to the right with probability $|\beta|^2$. 
An ordered time evolution of the QW is thus governed by unitary transformations that, starting from  the initial state,  transform it after $n$ steps into $\ket{\Psi(n)} = \hat{U}^n \ket{\Psi}$, being $\hat{U}$ a unitary operator. 
This comes in the form  $\hat{U} = \hat{S}\cdot (\hat{C} \otimes \hat{I})$, corresponding indeed to the  application of the coin flip operator $\hat{C}$, then followed by the shift operator $\hat{S} = \sum_{x} \ketbra{L}{L} \otimes \ketbra{x-1}{x} + \ketbra{R}{R} \otimes \ketbra{x+1}{x}$, which displaces the particle in a superposition of the position basis states.

This DTQW can be physically realized by sending photons through a cascade of directional couplers acting as symmetric beam splitters (BSs) arranged in a lattice of basic cells, each forming a Mach–Zehnder (MZ) interferometer (Fig. \ref{fig:cascade} top panel). In this scheme, each vertical row of couplers corresponds to a discrete time step, while horizontal lines represent the position state $\ket{x}$ of the walker. Hence, every BS output corresponds to a point in the space-time evolution of the QW. Each BS simultaneously implements both the coin and shift operators, since it sets the direction of the particle (left or right) and shifts it accordingly. When a photon at the position $\ket{x}$ enters the BS from the left or right side at the time step $n$ , its state can be written as $\ket{L,x}_n$ or $\ket{R,x}_n$, respectively.  For a balanced BS, the coin operator is in the Hadamard form $\hat{C} = \frac{1}{\sqrt{2}} \left( \begin{matrix} 1 && 1 \\ 1&&-1 \end{matrix} \right)$, thus the transition yields the states $\frac{1}{\sqrt{2}}(\ket{R,x+1}_{n+1} + \ket{L,x-1}_{n+1})$ or  $\frac{1}{\sqrt{2}}(\ket{R,x+1}_{n+1} - \ket{L,x-1}_{n+1})$.  Building on this framework, we consider evolution in a bidimensional discrete lattice, where the horizontal $n$-axis indicates the time steps ($0 \leq n \leq N, n \in \mathds{N}$) up to its maximum value $N$, and the vertical $x$-axis denotes the spatial sites accessible to the walker ($-M \leq x \leq M, M \in \mathds{N}$). In this scheme, $2M$ identifies the number of sites in the lattice at a given time, and $M$ is the maximum number of MZ units per time step. 

Disorder can be introduced in the form of random phase shifts between the MZ interferometers paths. In our model, we build at each time step a phase matrix $M_{\phi}(n)$ as a diagonal matrix with elements $p_{jj}(n)=e^{i \phi_{j,n}}, 1 \leq j \leq 2M$, where the phases $\phi_{j,n}$ are random values drawn from a uniform probability distribution over the interval $[p_{min}, p_{max}]$. Thereby the effect of the dynamic disorder is to introduce errors in the rotation of the qubit, due to the fact that coin operators have a random component. The noise-free configuration is recovered by setting $p_{min}=p_{max}=0$. A $2M\times2M$ block-diagonal matrix $M_{BS}$ at step $n$ is then built, where each diagonal block consists of a $2\times2$ BS transformation with reflectivity $r^2=1/2$, in order to implement a Hadamard coin. The block matrix structure properly accounts for the alternate pattern of an even and an odd number of BSs over subsequent time steps. At each time step, a different evolution matrix is computed as $U(n)=M_{BS} \cdot M_{\phi} (n)$, thus including the noise  introduced by random phases.
The complete evolution of the QW after $n$ steps is then obtained as the time-ordered product of unitary matrices representing each layer of BSs: $$\ket{\Psi(n)} = \hat{U}(n) \, \hat{U}(n-1) \ldots \hat{U}(2) \, \hat{U}(1)\ket{\Psi(0)}$$

In our investigation, noise is implemented by assigning the same coin operator to all spatial sites,  allowing it to vary at each time step. Specifically,  the same phase shifts are applied to all MZs belonging to a given step of the walk ($\phi_{x,n}=\phi_n, \forall x$).  As a result, we obtained a disorder varying in time, but independent on the walker position. We distinguish between two noise regimes  -- either \textit{strong} or \textit{weak}-- defined by the width of the uniform probability distribution from which the random phase values are drawn. In particular, strong noise corresponds to picking random phases over the full interval $[-\pi, \pi]$, while weak noise corresponds to random phases from the narrower interval $\left[-\frac{\pi}{8}, \frac{\pi}{8}\right]$. Two distinct scenarios are investigated: a \textit{time-unsorted} configuration, where dynamic noise varies randomly over time, and a \textit{time-sorted} configuration, where the dynamic noise remains random, but its intensity progressively increases with time, i.e. the steps of the QW.  

In a QW, the probability distribution of finding a walker at a given position after several steps exhibits distinctive interference patterns due to the coherent superposition of the many possible paths. To capture this behaviour, we look at the position probability distribution $P(x;n) = \sum_{k=R,L}|\langle k, x|\Psi(n)\rangle|^2$, which represents the probability of finding the particle, after $n$ steps, at position $x$, irrespective of its  coin state. In order to capture salient features of our confined walks without being sidetracked by specific cases, we adopt a single-particle probability distribution after $n$ time steps $P_M(x;n) = \frac{P(x;n)_a + P(x;n)_b}{2}$, averaged over two initial conditions labeled as $a$ and $b$ (Fig. \ref{fig:cascade} bottom panel). 

We analyse the propagation dynamics of a two-mode state under symmetric initial conditions, where the walker is injected into the lattice in a balanced superposition of the central sites of the position axis such that $x_M(n=0)=0$, where $x_M$ is the two photon mean position. These two configurations thus start at positions $\vert x_a\rangle$ and $\vert x_b\rangle=\vert -x_a\rangle$, and the coin is initialised in $\vert L \rangle$ and $\vert R \rangle$, respectively.  A quantitative indication on the spreading of the walker wavefunction is provided by the variance of this mean position as a function of the number of steps, defined as $\sigma^2(n) = \langle P_M(x;n)^2 \rangle-\langle P_M(x;n)\rangle^2$.

{\section{Simulations}\label{sec:simulations}}

\subsection{QWs in unbounded lattices}

\begin{figure}
    \centering     
    \includegraphics[width=1.\linewidth]{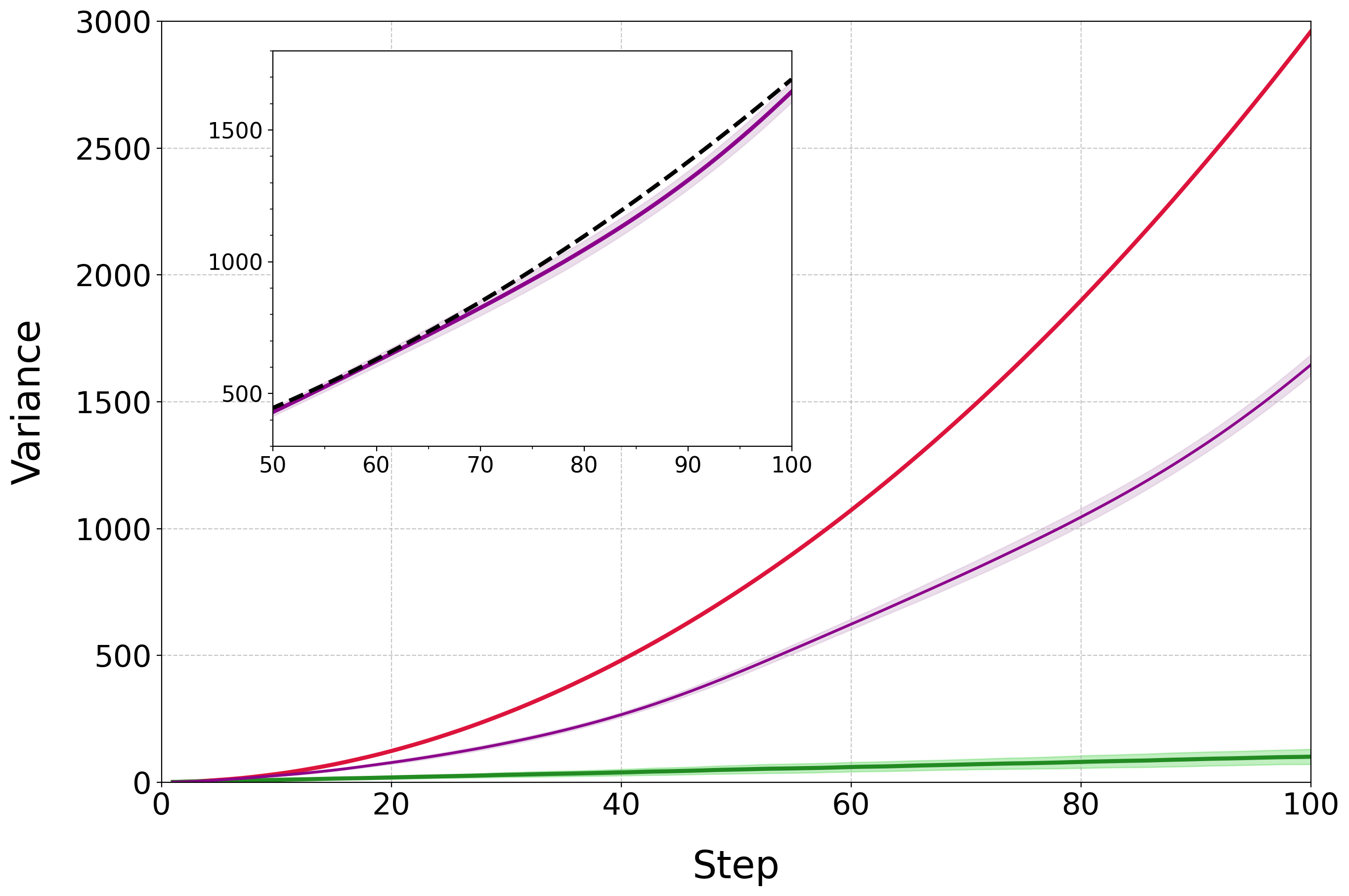}
     \caption{Time evolution of variance of the walkers' mean position (calculated with respect to the middle point of the spatial axis) in three different scenarios for QW in unbounded lattice: noise-free configuration (red line), QW with dynamic noise (green line), and QW with time-sorted dynamic noise (purple line). The dashed line in the inset indicates the quadratic trend followed by the variance when disorder increases with the time steps: after the first $\sim60-70$ steps, the dynamic of the QW starts to deviate from the ballistic regime, which is typically observed in the absence of noise. Probability distributions are averaged over 100 random configurations for each kind of noisy QW. The standard deviation of data points are represented by colored bands.   
    }
    \label{fig:Variance_open_QW}
\end{figure}

We first remind the behaviour of noisy QWs in arrangements without confinement. This implies that the transverse dimension of the lattice is virtually infinite and translational-invariant -- or, at least, large to the point its edges are never effectively populated, since the number of available sites $2M$ is never exceeded after the maximum number of steps $N$. This clearly implies that the walker can evolve in its spatial degree of freedom without encountering boundary effects up to the final step. In our simulation, we set $2M=200$, and $N=100$, in order to also capture long-time behaviours. The ordered configuration is retrieved by setting the local phase parameter as a constant, i.e., $\phi_{x,n}=0, \forall x \in [-M,M], \forall n \in [1,N]$. The time evolution of the walker is thus homogeneous. The result is the well-known quadratic increase of the variance with the time step, as shown in Fig.~\ref{fig:Variance_open_QW} (red line), showing a ballistic spread of the dynamics.

The time-translation symmetry of the quantum walk can then be broken by adding disorder. This amounts to allowing for an interaction between the quantum system and an external classical environment. In the presence of dynamic time-unsorted disorder, irrespective of the noise strength, the variance increases linearly with the number of steps, which is the hallmark of diffusive dynamics (Fig.~\ref{fig:Variance_open_QW}, green line) . In this regime, the photon propagation process reduces to a classical random walk. This is also supported by theoretical studies extended to $n$-dimensional QW \cite{Mackay2002,Kendon2007}, confirming that by introducing random phases and averaging over a large number of configurations, decoherence is established. In the time-sorted case, a more nuanced behaviour is observed. Unlike the time-unsorted case, the ballistic increase of the variance is not suppressed, but slowed (Fig.\ref{fig:Variance_open_QW}, purple line).

\subsection{Confined QW}
When the number of sites $2M$ is not large with respect to the scale of the diffusion, boundary conditions become significant. We simulate this dynamics by imposing a hard confinement in the lattice, in that the walker is completely reflected at the edges, at any step that would bring it outside. This confined QW is simulated on $2M=8$ sites over $N=100$ steps. 

\begin{figure}
    \centering        
    \includegraphics[width=\columnwidth]{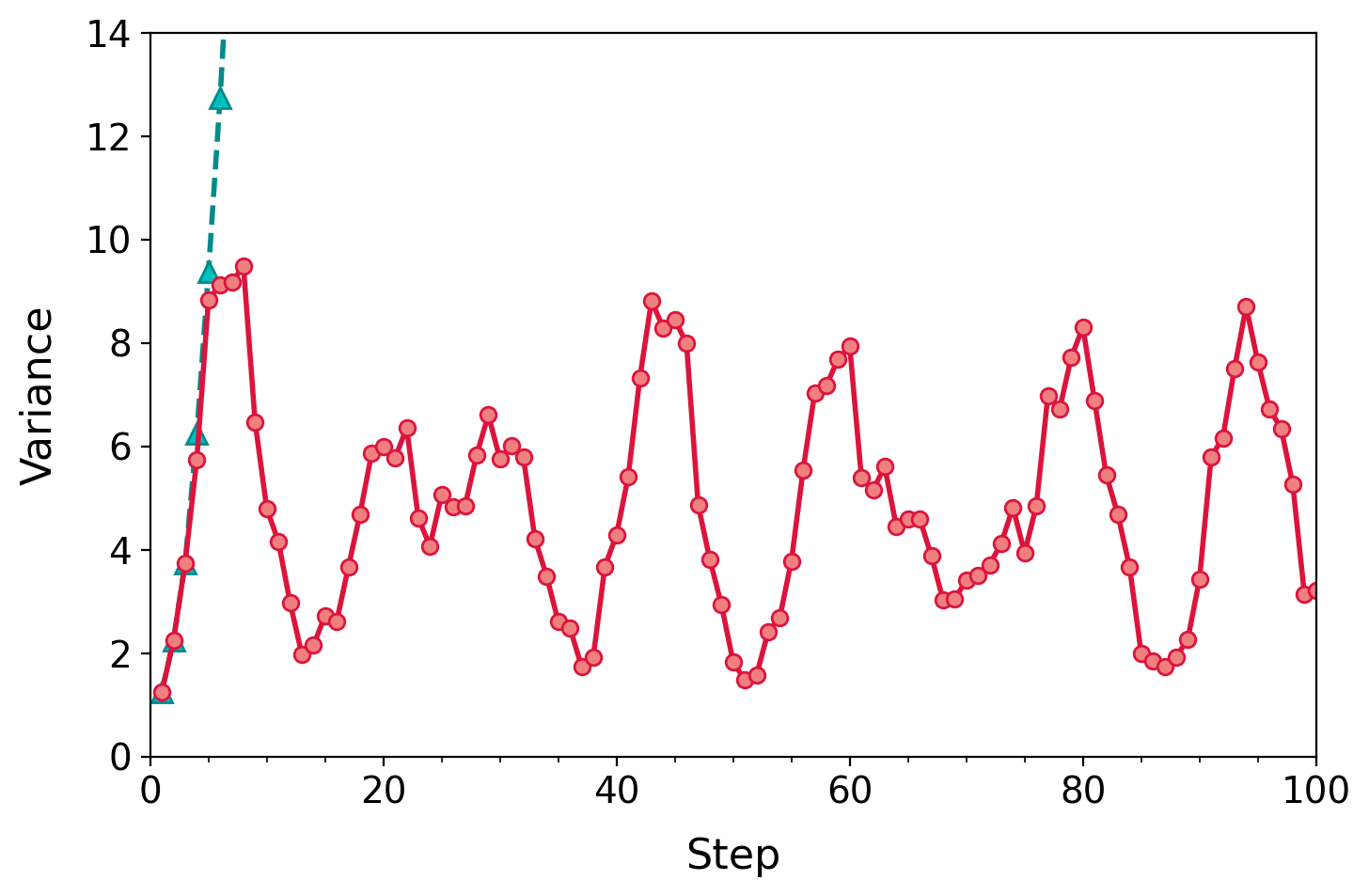}
     \caption{Time evolution of the variance of the mean position for a disorder-free QW: comparison between the propagation of walkers in confined (red circles) and unbounded (dark cyan triangles) lattice. Colored lines are shown as guides to the eye.}
     \label{fig:ConfinedQW_NoiseFree_Variance}
\end{figure}

\begin{center}
\begin{figure}[b]
\subfigure[]{
\includegraphics[width=\columnwidth]{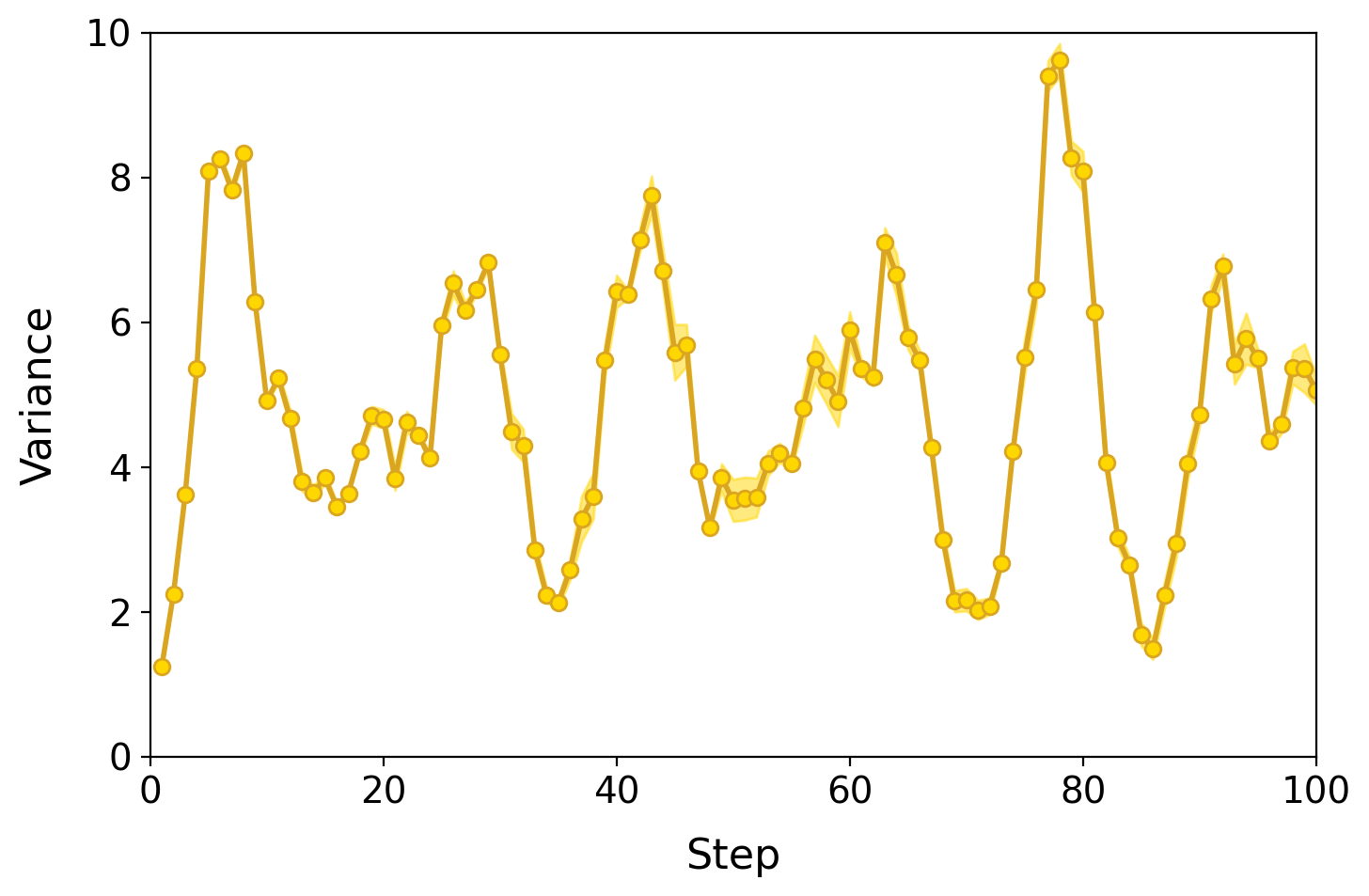}
}
\subfigure[]{
\includegraphics[width=\columnwidth]{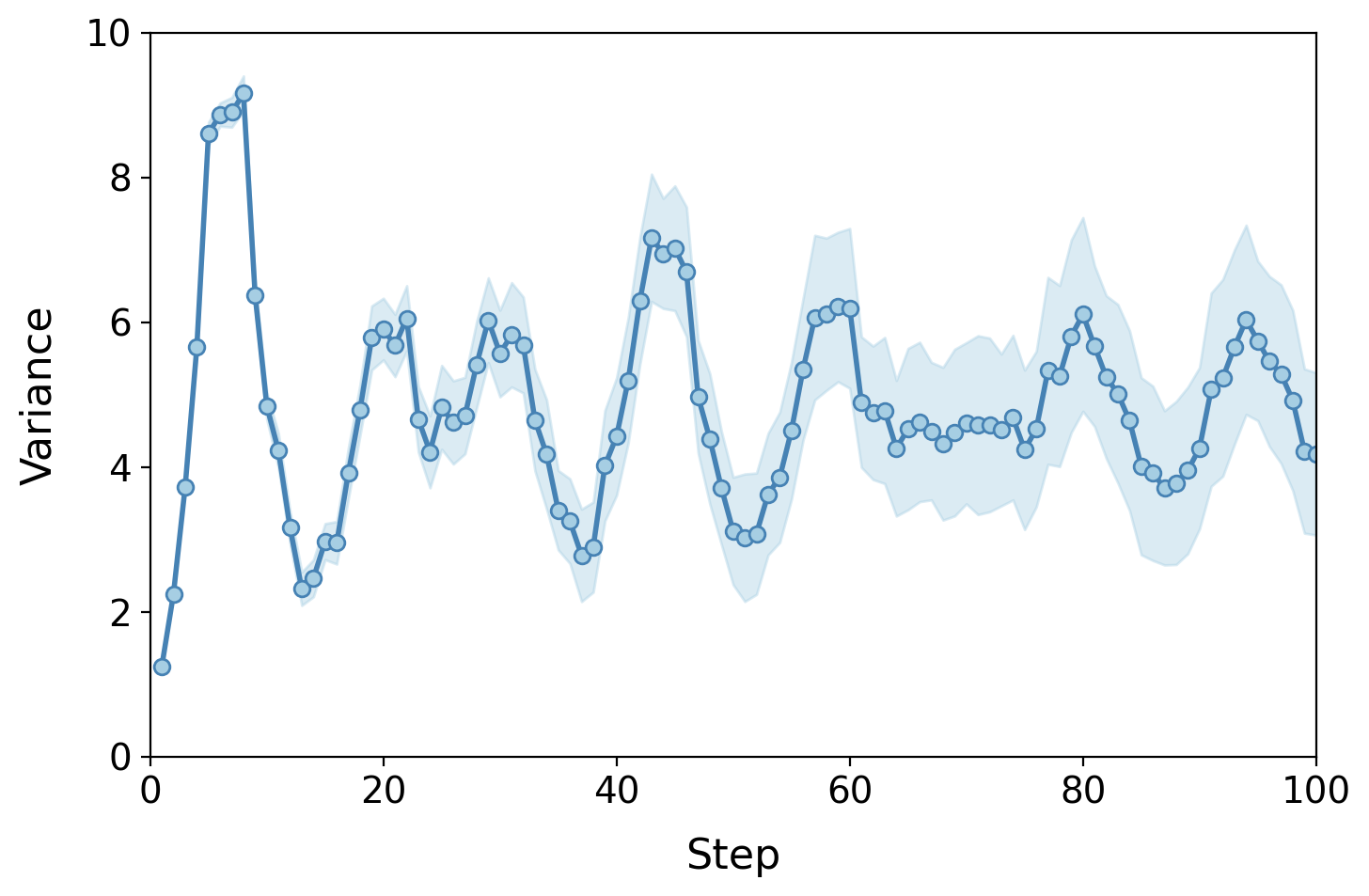}
}
\caption{Variance of the mean position as a function of the number of steps for a confined QW with dynamic weak noise (colored lines are shown as guides to the eye). Probability distributions are averaged over 100 random configurations for each kind of noisy QW. Errors are shown as colored bands. (a) Dynamic noise with stepwise increasing phases (time-sorted disorder). (b) Dynamic time-unsorted noise.}
\label{fig:confinedQW_dynamic_weak}
\end{figure}
\end{center}

\begin{center}
\begin{figure}[h]
\subfigure[]{
\includegraphics[width=\columnwidth]{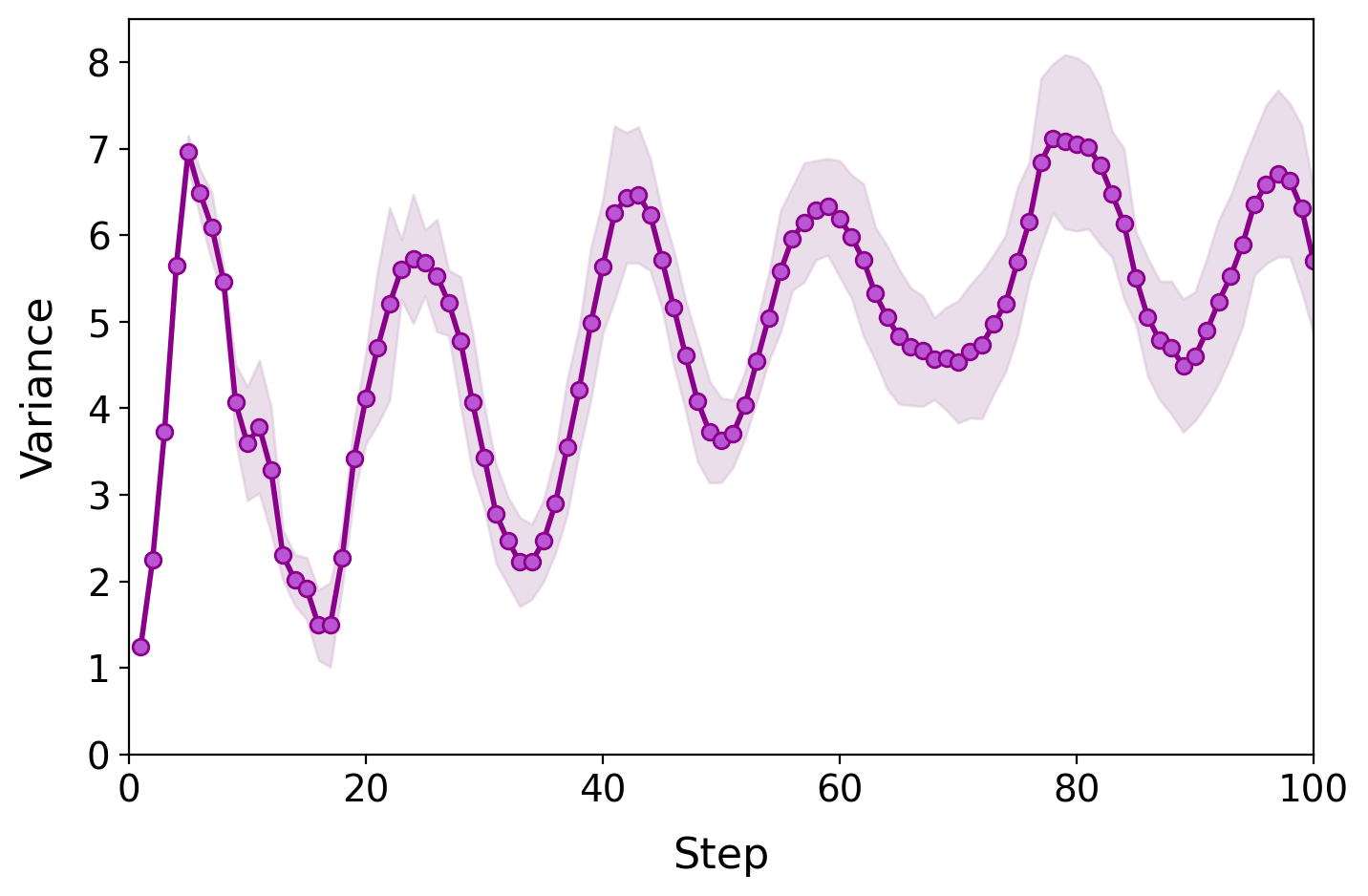}
}
\subfigure[]{
\includegraphics[width=\columnwidth]{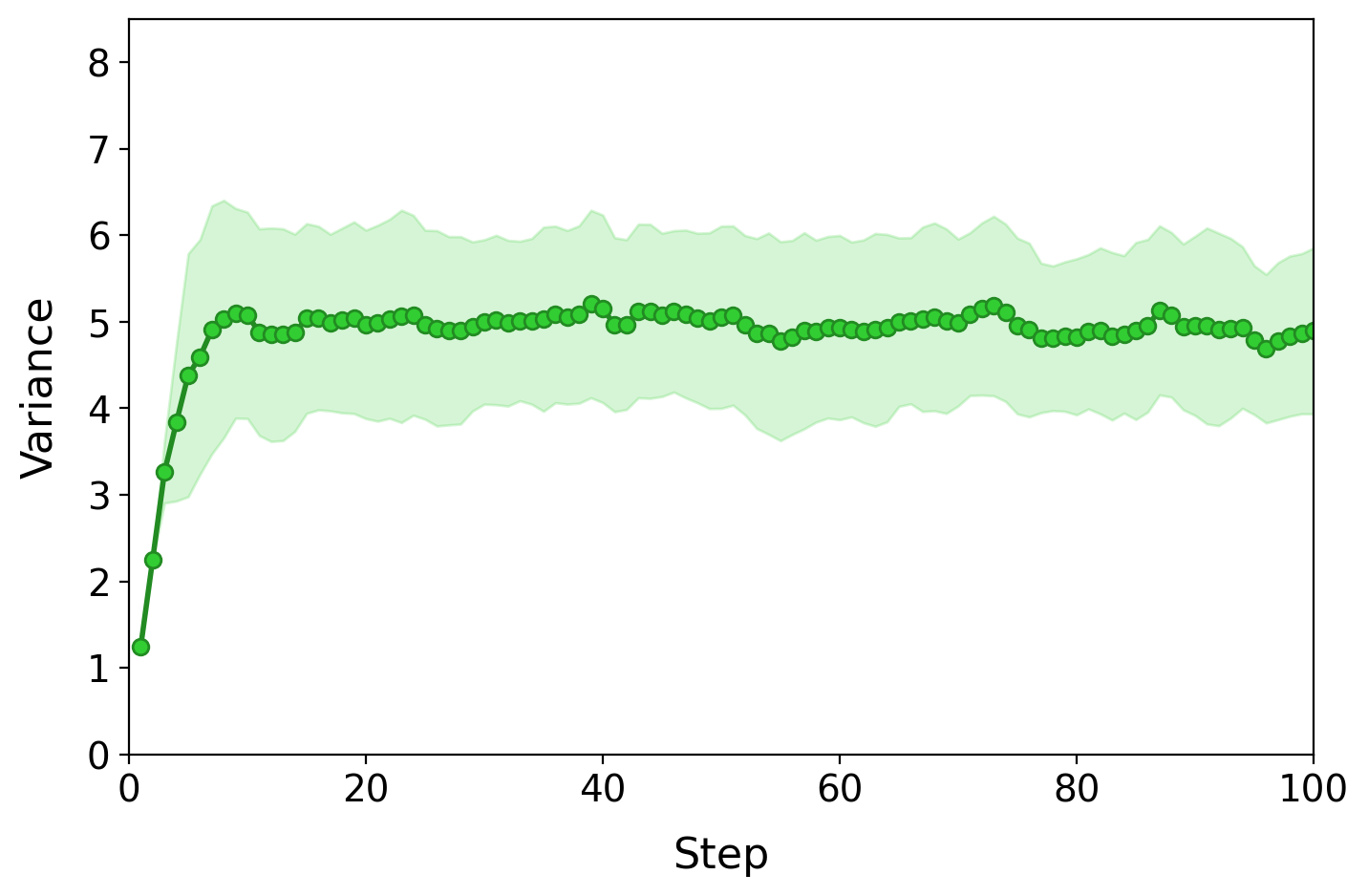}
}
\caption{Variance of the mean position as a function of the number of steps for a confined QW with dynamic strong noise (colored lines are shown as guides to the eye). Probability distributions are averaged over 100 random configurations for each kind of noisy QW. Errors are shown as colored bands. (a) Dynamic noise with stepwise increasing phases (time-sorted disorder). (b) Dynamic time-unsorted noise.}
\label{fig:confinedQW_dynamic_strong}
\end{figure}
\end{center}

\begin{figure}[h]
    \centering
\includegraphics[width=0.9\columnwidth]{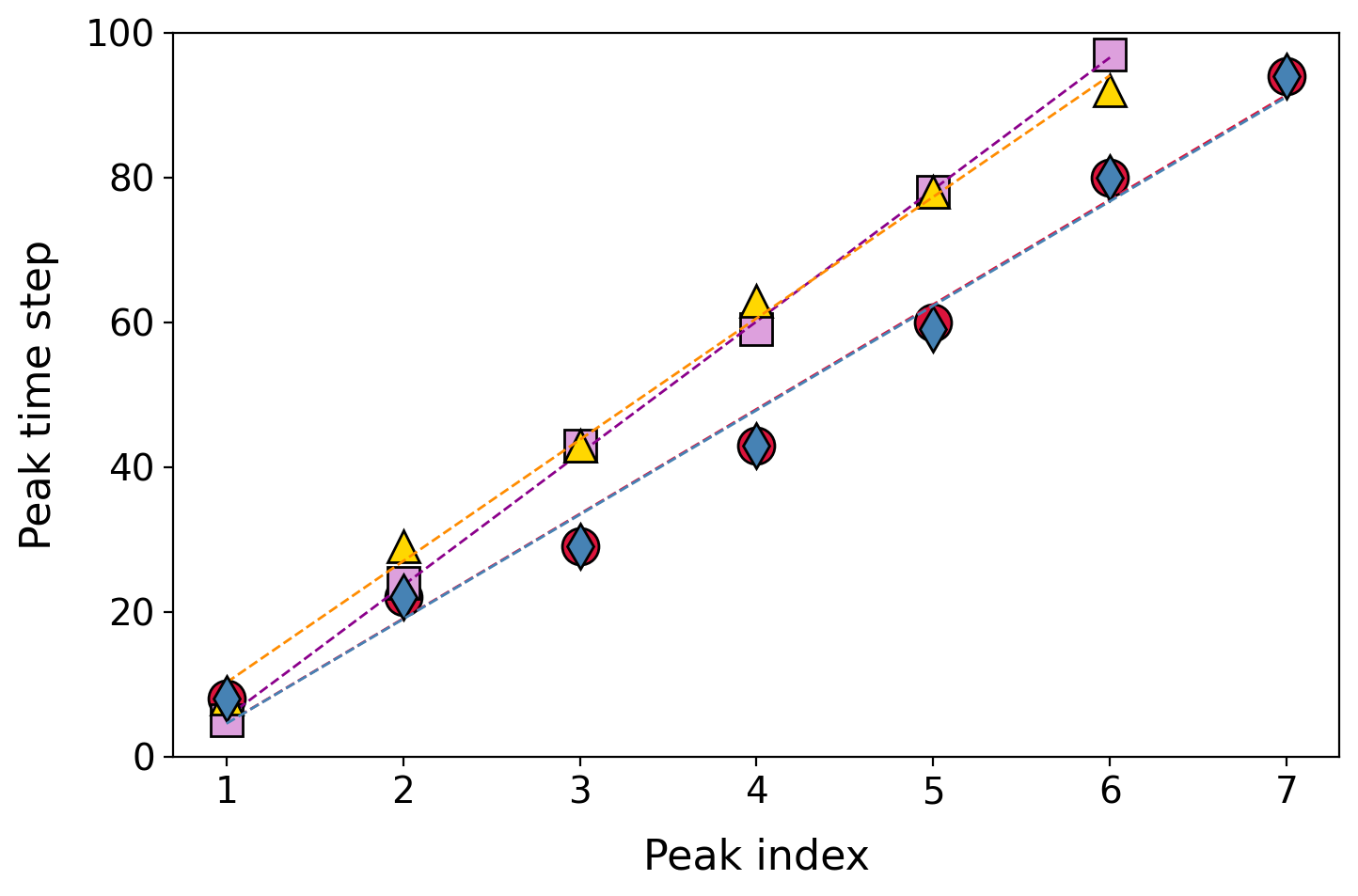}
    \caption{Steps at which peaks of the mean position variance occur as a function of their positional order within the sequence of relative maxima, for confined QW without disorder (red circles) and for confined QW subject to strong time-sorted (pink squares), weak time-sorted (yellow triangles), and weak time-unsorted (blue diamonds) dynamic noise. Data are referred to the variance averaged over 100 random configurations given in Fig. \ref{fig:confinedQW_dynamic_weak} and Fig.\ref{fig:confinedQW_dynamic_strong} (a). 
    The linear trends (dotted lines) indicate a recursive behavior in the variance dynamics, which is faster without temporal phase sorting, as reflected by the smaller slope of the fitted line.
    }
    \label{fig:Peaks_comparison}
\end{figure}

In the absence of disorder, the variance of the walker's mean position exhibits an oscillatory behavior as a function of the number of steps, alternating spreading and shrinking of the wavefunction, as in Fig. \ref{fig:ConfinedQW_NoiseFree_Variance}. This oscillatory is in stark contrast with the monotonic increase of the unconfined case. Such a trend emerges as the result of interference effects induced by boundary reflections. These bring about the generation of standing-wave–like patterns within the confined region, although there is no overall pattern replicating indefinitely.
We then consider the introduction of dynamical noise. The case of weak noise is illustrated in Fig. \ref{fig:confinedQW_dynamic_weak}. The variance of the walker mean position as a function of the time steps exhibits a similar oscillatory behaviour in both time-sorted and time-unsorted regimes. As a remarkable difference, the amplitude of the oscillations is reduced in the time-unsorted configuration at long times, since transport is hindered. 
Fig.~\ref{fig:confinedQW_dynamic_strong} reports on the strong noise condition, with the time-sorted and time-unsorted regimes exhibiting markedly different behaviours. In the former case, the variance of the walker  position maintains oscillations similar to the previous case, with a tendency to increase at long time. In contrast, in the time-unsorted scenario, the variance rapidly saturates within the first $\sim10$ steps, showing only small residual oscillations around a mean value comparable to that obtained in the other instances. 
Beyond qualitative similarities, a more detailed analysis reveals that time-sorted noise affects the time scale of the oscillations regardless its intensity. Fig. \ref{fig:Peaks_comparison} reports the steps at which the variance hits a peak in its oscillations. Time sorting appears to slow the dynamical evolution on average, with comparable results in the weak and strong regimes.

In summary, in the presence of dynamic noise, it is possible to induce a controlled transition of the system toward a more localized dynamics either by introducing strong disorder and scrambling phases over time or by adding a time-unsorted noise and increasing its strength.  

\section{Experimental confined QWs\label{sec:experiment}}

\subsection{The photonic platform}
The behaviour of a confined quantum walker has been tested using a reconfigurable photonic processor (QuiX Quantum Alquor20) with 20 input and output ports \cite{Taballione2021,Taballione2023}. This is an programmable multiport interferometer, capable of  implementing arbitrary  linear optical transformations on a space as large as the number of input/output modes.  The multiport interferometer itself is built as a mesh of 190 Mach-Zehnder interferometers (MZIs), each acting as a tunable beam splitter~\cite{Clements16}, thus ensuring  control both amplitudes and phases of its two output signals. The photonic chip is based on stoichiometric silicon nitride $Si_3N_4$ waveguides using TripleX technology and is mounted on a water-cooled Peltier element in order to maintain a constant temperature. Optical access to and from the chip is ensured by means of FC/PC fibre connectors. The insertion loss of the processor  has been measured to be $(3.65 \pm 1.30)\ dB$ and the average amplitude fidelity over 100-Haar random matrices is measured to be F = $(98.8 \pm 0.3) \%$ at 942 nm where F = $\frac{1}{d} Tr(|U^{\dagger}_{th} \cdot U_{exp}|)$, $|U|$ is the component-wise absolute value and $d = 20$ is the number of ports~\cite{wang09}.  The fully packaged photonic chip sits on a sub-mount for mechanical stability and is electronically connected to a printed circuit board (PCB). The user interface is realised by a software in Python allowing to input the desired transformation, so as to implement the necessary phase shifts in the MZIs. 

\begin{figure}[b]
    \centering
    \includegraphics[width=0.95\linewidth]{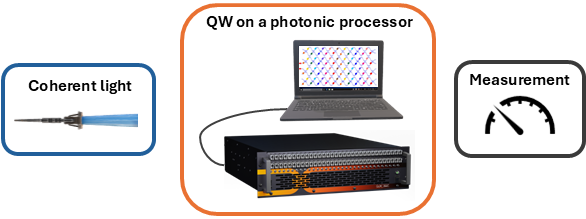}
    \caption{Conceptual scheme of the experimental setup. Attenuated coherent light plays the role of the walker, the tunable photonic processor implements the QW unitary and a photodiode is used for the measurement of probability distribution across the output modes. Thanks to a PC driven reconfigurability of the processor, we implemented various noise configurations.}
    \label{fig:setup}
\end{figure}
A scheme of the experimental setup is reported in Fig. \ref{fig:setup}. Attenuated coherent light was injected through polarization-maintaining fibers in the desired input modes of the processor which implements a unitary according to the confined QW and disorder regime of interest. We found that elaborating the form of the transformation off-line and using the proprietary software for implementation was more accurate than attempting to introduce noise by setting individual phases. This is due to the fact the original software also corrects for cross-talk across different MZIs. The size of the multimode interferometer allowed us to to implement up to $N=20$ steps of the quantum walk; in the transverse dimension we have used $2M=8$ input and output ports. Input modes $4$  for $x_a$ and 5 for $x_b$ in the selected set were initially populated by injecting $\approx 10\, \mu W$ using polarization-maintaining fibers.  After applying the proper unitary operator for an $n$-step quantum walk, with $4\leq n\leq 20$, we measured output intensities on all 8 output channels by means of photodiodes reached via single-mode fibers, thus retrieving the output probability distributions. Since the output fibres may have different transmissions, we have calibrated these by using a fixed output when applying the identity on the processor. Following this correction, the distributions were normalised to unity.
\subsection{Results\label{sec:results}}
\begin{figure}[t]
    \centering
    \includegraphics[width=0.9\columnwidth]{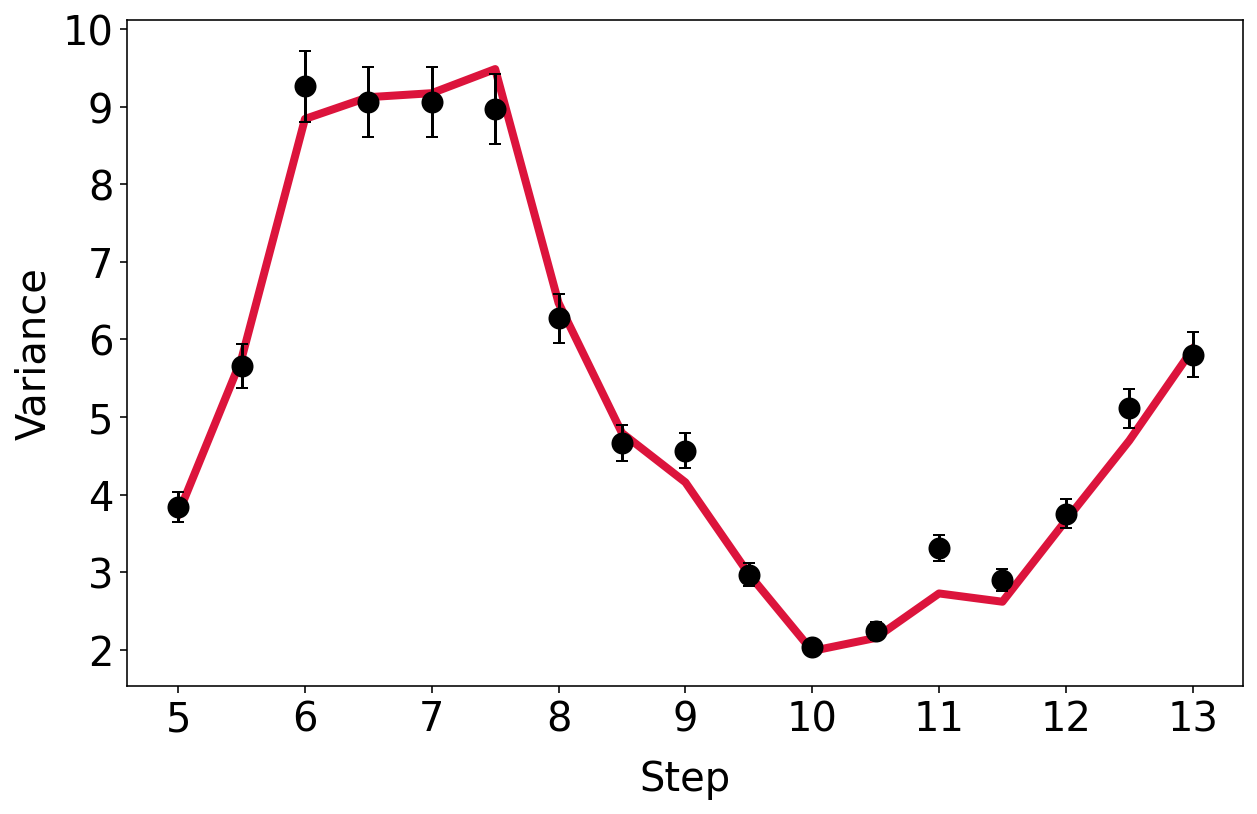}
    \caption{Variance of the mean position in a noise-free 4-site QW as a function of the number of steps. Dots represent experimental data, the line shows the expected behavior.
    }
    \label{fig:QW ordered variance}
\end{figure}
First, a confined quantum walk was implemented in the absence of noise. As outlined in the previous Sec.~\ref{sec:simulations}, under these conditions, the wavefunction is expected to exhibit recursive
spreading and refocusing throughout its evolution. This dynamic behavior is manifested as oscillations in the variance of the mean position. The experimental results for the variance are presented in Fig.~\ref{fig:QW ordered variance} and reflect the expected behaviour: the experimental data is shown as dark circles, and the numerical simulation as the red line, with remarkable agreement. 
Appendix \ref{app:distributions} present full plots of the recorded output probability distributions at all time steps.

As a second step, we investigated how sorted dynamic noise influences the confined quantum walk, accessing, to this end, both weak and strong regimes.  From the ensemble of phase configurations used in the simulations in Sec. \ref{sec:simulations}, we selected a representative set and implemented it on the photonic processor (for the corresponding phase vectors, see Table \ref{tab:Sorted_phases} in Appendix \ref{app:parameters}).
\onecolumngrid
\begin{figure*}[t]
    \centering 
        \includegraphics[width=\textwidth]{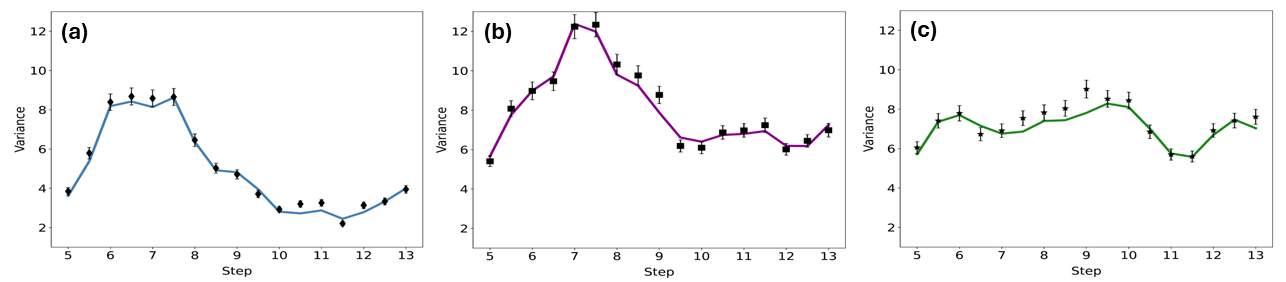}
     \caption{Variance of the mean position for a 4-site QW with (a) weak, (b) strong sorted and (c) strong unsorted dynamic noise as a function of the number of steps. Dots represent experimental data, lines show the expected behavior.
     }
    \label{fig:QW_dynamicNoise_Sorted}
\end{figure*}

\twocolumngrid
The measured variances of the mean position for the two noise regimes are presented in Fig. \ref{fig:QW_dynamicNoise_Sorted} (a) and (b). The graphs reveal excellent agreement between the experimental data and the numerical simulations. A comparison between the two noise regimes shows a markedly larger spreading in the presence of strong sorted dynamic noise, highlighted by the higher peak of the variance. This regime also exhibits a faster temporal evolution: the wavefunction shrinks more rapidly, reaching its minimum variance after 12 steps, whereas in the weak-noise regime the corresponding minimum occurs only after 17 steps. Although the dynamics is faster, the minimum variance in the presence of unsorted noise is higher than in previous cases, suggesting a breakdown in what would otherwise appear to be a recurring pattern (see Fig. \ref{fig:maps} in Appendix \ref{app:distributions}).\\
In the final set of experiments, we explored the effects of unsorted dynamic noise. As discussed in Sec. \ref{sec:simulations}, the distinctive features of unsorted noise emerge predominantly in the strong-noise regime; accordingly, we limited our investigation to this configuration. For this measurement, we selected a representative phase vector from the ensemble used in the simulations (see values in Table \ref{tab:Sorted_phases} of SI, corresponding to $\phi_{\mathrm{Unsorted}}^S$) .
The output distributions were measured and the corresponding variance was computed as a function of the number of steps. The results, shown in Fig. \ref{fig:QW_dynamicNoise_Sorted} (c), exhibit good agreement between experiment and theory. The oscillatory behavior seen in previous cases is notably reduced, revealing that unsorted dynamic noise causes decoherence and weakens the interference effects of the Hadamard walk.

\section{Conclusions}
\label{sec:conclusions}
We presented a study on confined {discrete-time} quantum walks where the number of steps exceeds the number of accessible lattice sites, focusing on how the walker evolves under dynamic noise. Simulations on an 8-site lattice over 100 steps reveal that confinement induces an oscillatory—though not strictly periodic—spreading of the walker’s wavefunction, mirrored in the behavior of the position variance.
We further demonstrate that dynamic disorder, whether weak or strong, {does not act on the dynamics merely as a detrimental perturbation, but also} significantly influences both the overall spreading and the effective propagation velocity. Notably, even when the noise grows in time, a meaningful degree of coherence can still be maintained provided that the underlying noise level remains low. Our results indicate that a key parameter governing this resilience is the magnitude of the dynamic noise itself, beyond its temporal {structure} during the evolution.
These findings enable us to identify a regime in which the testbed protocol becomes more robust to noise, exhibiting dynamics that closely resemble those expected in an ideal, noise-free scenario.
This fault-tolerant regime could be especially relevant for emerging applications in quantum information processing, particularly as integrated photonic platforms continue to scale in size and complexity \cite{EuorHPC,Wang2020,Epique,Carolan15,Quix,EPHOS}.\\\\

\noindent\textbf{Acknowledgments:} This work was funded by QuantERA II Programme supported by the EU H2020 research and innovation programme under GA No 101017733, with funding Italian organization PNRR MUR project
PE0000023-NQSTI (Spoke 6, CUP: H43C22000870001). IG and CB are supported by the PRIN 2022 MUR Project EQWALITY (N. 202224BTFZ). MB is supported by Rome Technopole Innovation Ecosystem (PNRR grant M4-C2-Inv). IG and MB acknowledge the support from MUR Dipartimento di Eccellenza 2023-2027.\\
\textbf{Availability of data and materials} The datasets used and analyzed during the current study are available from the corresponding author on reasonable request.\\
\textbf{Conflict of interest} The authors declare no conflicts to disclose.

\bibliographystyle{quantum.bst}
\bibliography{Bibliography_Confined_QW}

\newpage
\onecolumngrid
\appendix
\section{Experimental probability distributions}
\label{app:distributions}
\begin{figure}[h]
    \includegraphics[width=\linewidth]{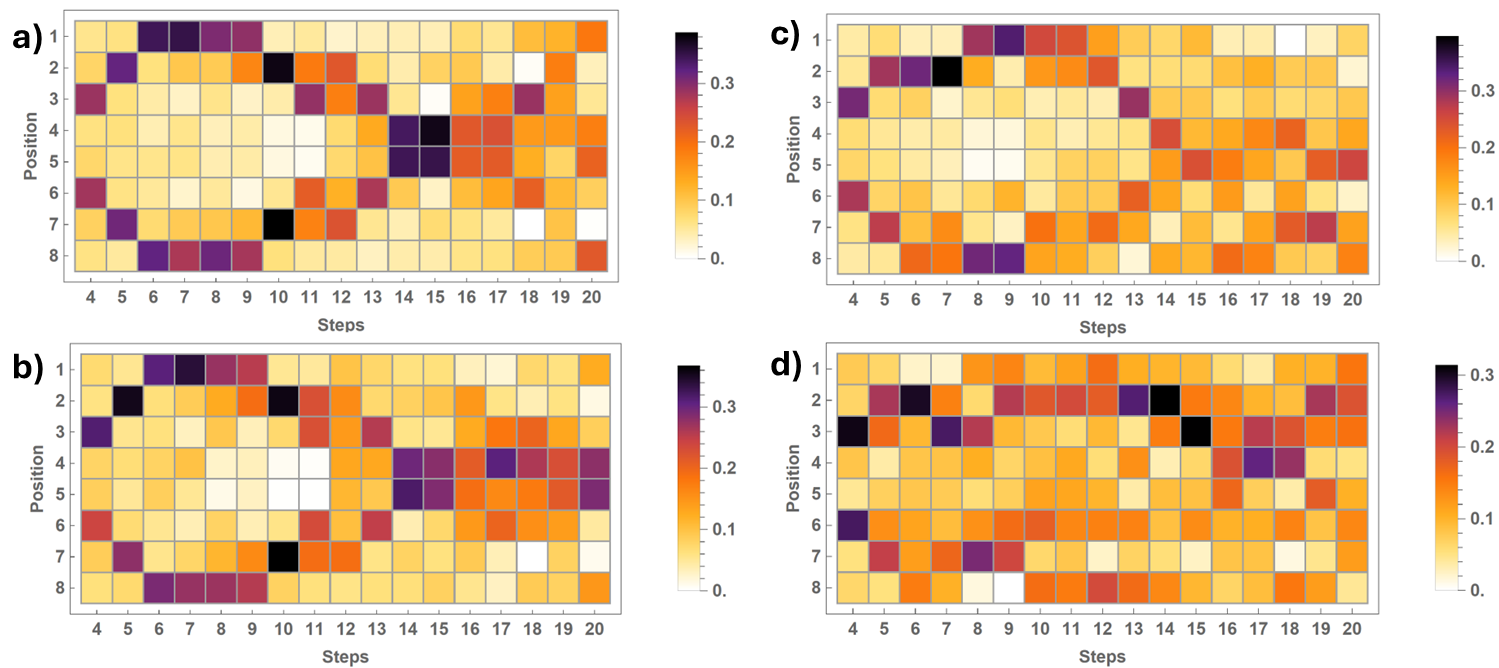} 
    \caption{Experimental probability distributions of a confined 4-sites quantum walk from 4 to 20 steps in the case of a) Hadamard walk, b) in the presence of dynamic sorted weak noise, c) dynamic sorted strong noise, d) dynamic unsorted noise.}
    \label{fig:maps}
\end{figure}

As detailed in the main text, we measured the output probability distributions of the quantum walker in a confined lattice after a series of steps. Figure \ref{fig:maps} presents a color map of these distributions, showing the results from our experiments both in the absence of noise and under the influence of dynamic noise across the different regimes outlined in the main text.

To quantitatively assess the agreement between the experimental and theoretical probability distributions, we computed the total variation distance (TVD), defined as the maximum absolute difference between the theoretical and experimental output probabilities:
\begin{equation}
\mathrm{TVD} = \max_i \left| P_i^{\mathrm{theo}} - P_i^{\mathrm{exp}} \right|.
\end{equation}

\begin{figure}[h]
    \includegraphics[width=0.6\linewidth]{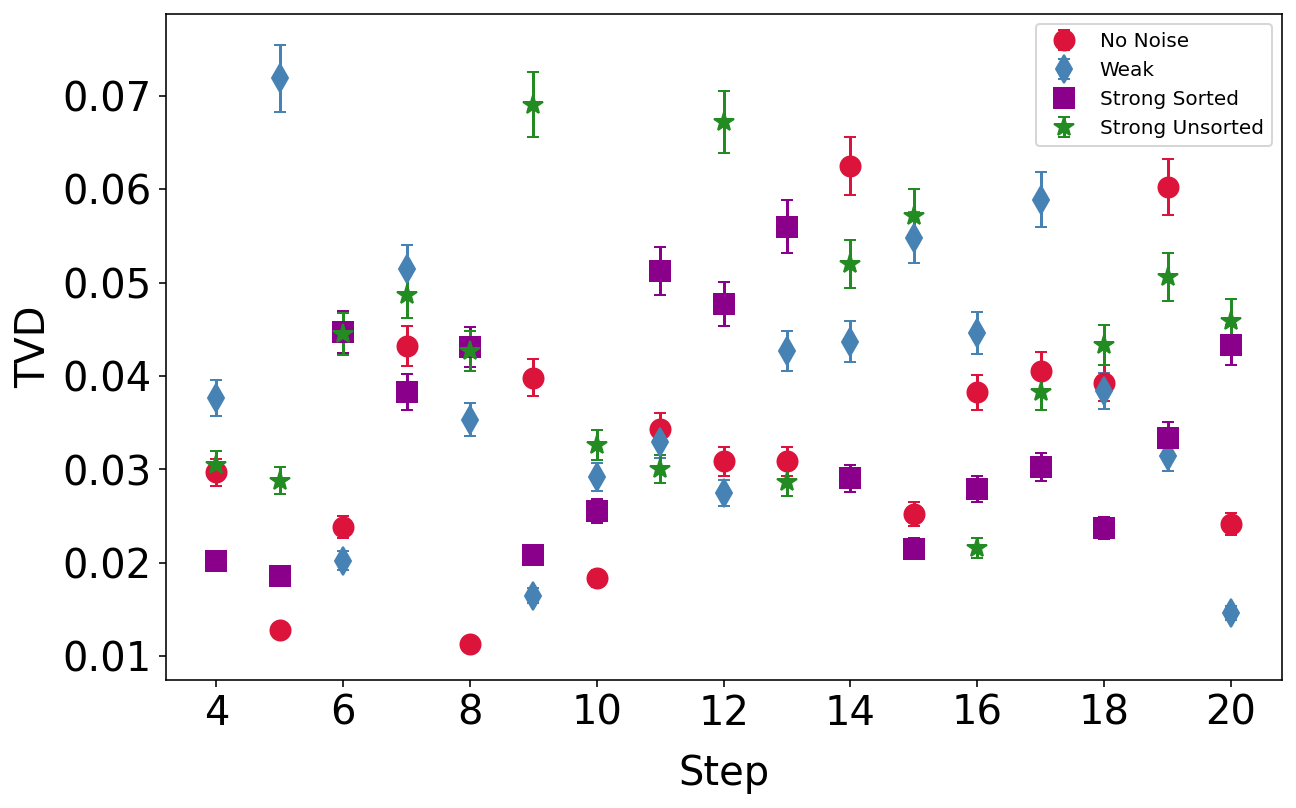} 
    \caption{TVD of the experimental probability distributions for a 8 sites up to 20 steps quantum walk in the case of a hadamard walk (red dots), in the presence of dynamic sorted weak (cyan diamonds), sorted strong (violet squares) and unsorted (green stars) noise.}
    \label{fig:tvd}
\end{figure}
A smaller TVD indicates a closer match between the experimental results and the theoretical predictions. Values of the TVD are reported in Fig. \ref{fig:tvd}. We can clearly observe that the values are fairly close to zero, and all of them below 0.8, highlighting the good agreement between experimental and simulated distributions

\section{Experimental parameters}
\label{app:parameters}
As reported in the main text, we implemented specific sets of phases depending on the type of disorder we wanted for the walker to undergo. Specifically we report the values in Table \ref{tab:Sorted_phases}, where each row corresponds to dynamic sorted weak, sorted strong and unsorted noise.

\begin{table}[h]
    \centering
    \begin{tabular}{||c|c|c|c|c|c|c|c|c|c|c||}
    \hline\hline
         $n$ & 1 & 2 & 3 & 4 & 5 & 6 & 7 & 8 & 9 & 10\\\hline
         $\phi_{\mathrm{sorted}}^W$ & -0.387 & -0.380 & -0.368 & -0.285 & -0.275 & -0.270 & -0.261 & -0.260 & -0.244 & -0.140\\\hline
         $\phi_{\mathrm{sorted}}^S$ & -3.094 & -2.936 & -2.850 & -2.482 & -0.655 & -0.610 &-0.526 & -0.318 & -0.158 & 0.072\\\hline
         $\phi_{\mathrm{Unsorted}}^S$ & -0.156 & 2.983 & -2.675 & 1.609 & 0.636 & -1.676 & 1.763 & -1.572 & 1.623 & -2.235\\\hline\hline
         $n$ & 11 & 12 & 13 & 14 & 15 & 16 & 17 & 18 & 19 & 20\\\hline
         $\phi_{\mathrm{sorted}}^W$ & 0.016 & 0.019 & 0.117 & 0.198 & 0.221 & 0.242 & 0.243 & 0.353 & 0.374 & 0.392\\\hline
         $\phi_{\mathrm{sorted}}^S$ & 0.925 & 1.370 & 1.388 & 1.478 & 1.622 & 2.242 & 2.351 & 2.522 & 2.694 & 2.851\\\hline
         $\phi_{\mathrm{Unsorted}}^S$ & -1.240 & -2.981 & 1.767 & 0.274 & -1.402 & 0.850 & 2.797 & 1.779 & 1.332 & 2.775\\\hline\hline
    \end{tabular}
    \caption{Phases adopted at each step in the weak ($\phi_{\mathrm{sorted}}^W$) and strong ($\phi_{\mathrm{sorted}}^S$) sorted dynamic noise and unsorted ($\phi_{\mathrm{Unsorted}}^S$) dynamic noise regime.}
    \label{tab:Sorted_phases}
\end{table}

In this configuration, the $n^{\mathrm{th}}$ element of the table corresponds to the phase applied at the $n^{\mathrm{th}}$ step of the walk.

\twocolumngrid
\newpage

\end{document}